\documentclass[twocolumn]{aastex63}

\usepackage{color}
\usepackage{amsmath}
\usepackage{graphicx}

\shorttitle{Hydrogen-rich Atmospheres of the TRAPPIST-1 Planets}
\shortauthors{Hori \& Ogihara}

\begin{document}

\title{Do the TRAPPIST-1 Planets Have Hydrogen-rich Atmospheres?}

\correspondingauthor{Y. Hori}\email{yasunori.hori@nao.ac.jp}

\author[0000-0003-4676-0251]{Yasunori Hori}
\affiliation{Astrobiology Center, 2-21-1 Osawa, Mitaka, Tokyo 1818588, Japan}
\affiliation{National Astronomical Observatory of Japan, 2-21-1 Osawa, Mitaka, Tokyo 1818588, Japan}

\author[0000-0002-8300-7990]{Masahiro Ogihara}
\affiliation{National Astronomical Observatory of Japan, 2-21-1 Osawa, Mitaka, Tokyo 1818588, Japan}

\begin{abstract}
Recently, transmission spectroscopy in the atmospheres of the TRAPPIST-1 planets revealed flat and featureless absorption spectra, which rule out cloud-free, hydrogen-dominated atmospheres. Earth-sized planets orbiting TRAPPIST-1 likely have either a clear or a cloudy/hazy, hydrogen-poor atmosphere. In this paper, we investigate whether a proposed formation scenario is consistent with expected atmospheric compositions of the TRAPPIST-1 planets. We examine the amount of a hydrogen-rich gas that TRAPPIST-1-like planets accreted from the ambient disk until disk dispersal. Since TRAPPIST-1 planets are trapped into a resonant chain, we simulate disk gas accretion onto a migrating TRAPPIST-1-like planet. We find that the amount of an accreted hydrogen-rich gas is as small as 10$^{-2}$\,wt\% and 0.1\,wt\% for TRAPPIST-1\,b and 1\,c, 10$^{-2}$\,wt\% for 1\,d, 1\,wt\% for 1\,e, a few\,wt\% for 1\,f and 1\,g and 1\,wt\% for 1\,h, respectively. We also calculate the long-term thermal evolution of TRAPPIST-1-like planets after disk dissipation and estimate the mass loss of their hydrogen-rich atmospheres driven by a stellar X-ray and UV irradiation. We find that all the accreted hydrogen-rich atmospheres can be lost via hydrodynamic escape. Therefore, we conclude that TRAPPIST-1 planets should have no primordial hydrogen-rich gases but secondary atmospheres such as a Venus-like one and water vapor, if they currently retain atmospheres.
\end{abstract}


\section{Introduction}\label{sec:intro}

TRAPPIST-1 is an ultracool red dwarf with mass of $0.089\,M_\odot$ near the boundary between brown dwarfs and stars \citep{2018MNRAS.475.3577D,2018ApJ...853...30V}, located 12.43 pc\footnote{The parallax of TRAPPIST-1 from the {\it Gaia} mission: {\it Gaia} DR2 archive (https://gea.esac.esa.int/archive/).} away from Earth. Recently, seven Earth-sized planets orbiting TRAPPIST-1 were reported \citep{2016Natur.533..221G,2017Natur.542..456G}, three of which dwell in the conventional habitable zone. Transit timing variation (TTV) analyses of the TRAPPIST-1 planets find that their masses range from Mars to Earth \citep{2017Natur.542..456G,2017arXiv170404290W,2018A&A...613A..68G}. The TRAPPIST-1 planets are suited for studying atmospheric characterization and habitability of terrestrial planets beyond the solar system.

The low densities of the TRAPPIST-1 planets (see Table \ref{tab:trappist-1}) may conceal substantial volatile content, e.g., water inside their cores.
Monte Carlo studies on their internal compositions suggest that the inner planets have a small water content \citep{2017ApJ...842L...5Q,2018A&A...613A..68G}, which depends on their core-to-mantle mass ratio  \citep{2018ApJ...865...20D,2018RNAAS...2b..31S,2018NatAs...2..297U,2018RNAAS...2..116U}.
Interior modeling of the TRAPPIST-1 planets using updated masses indicates that the mass fractions of water might be uniform or increase with semimajor axis ($\lesssim$ 25\,wt\%) \citep{2018ApJ...865...20D}.
The current locations of the six inner TRAPPIST-1 planets are inside a snow line. Their volatile reservoirs should come from inward transport of building blocks such as icy pebbles \citep{2019A&A...624A..28I} or cometary material \citep{2018MNRAS.479.2649K} and planetary embryos that formed outside.
In fact, the six inner planets are trapped into a resonant chain; the outermost planet, TRAPPIST-1\,h, may form the Laplace relations \citep{2017NatAs...1E.129L}. In addition, transit geometry of the TRAPPIST-1 planets shows that they are in a nearly edge-on, co-planar system. 
The orbital architecture of the co-planar, resonant TRAPPIST-1 system supports that the planets experienced orbital migration \citep[e.g.][]{2016Natur.533..509M,2017A&A...604A...1O,2018MNRAS.476.5032P}. A disk migration scenario is explain the long-lived dynamical stability of the tightly packed TRAPPIST-1 system \citep{2012Icar..221..624M,2017ApJ...840L..19T}, which depends on the strength of tidal interactions between the inner planets and the central star \citep{2018MNRAS.476.5032P}.

\begin{table*}[]
    \centering
    \begin{tabular}{lcccc}
        \hline \hline
        Planet & Mass ($M_\oplus$) & Radius ($R_\oplus$) & Density ($\rho_\oplus$) & Semimajor Axis (au) \\
        \hline 
        b & 1.017 & 1.121 & 0.726 & 0.01154775 \\
        c & 1.156 & 1.095 & 0.883 & 0.01581512 \\
        d & 0.297 & 0.784 & 0.616 & 0.02228038 \\
        e & 0.772 & 0.910 & 1.024 & 0.02928285 \\
        f & 0.934 & 1.046 & 0.816 & 0.03853361 \\
        g & 1.148 & 1.148 & 0.759 & 0.04687692 \\
        h & 0.331 & 0.773 & 0.719 & 0.06193488 \\
         \hline
    \end{tabular}
    \caption{Observed physical properties of the TRAPPIST-1 planets \citep{2018A&A...613A..68G}}
    \label{tab:trappist-1}
\end{table*}

{\it Hubble Space Telescope} (HST)/Wide Field Camera\,3 (WFC3) measurements exhibit no prominent absorption features at near-infrared wavelengths in transmission spectra of six of the TRAPPIST-1 planets \citep{2016Natur.537...69D,2018NatAs...2..214D,2018AJ....156..178Z,2019MNRAS.487.1634B}\footnote{Stellar contamination effects on transmission spectra are expected to be less significant than those predicted in \citet{2018AJ....156..178Z} \citep{2018AJ....156..218D,2018ApJ...857...39M,2018ApJ...863L..32M}}. The combined spectrum of the planets rules out cloud-free, hydrogen-dominated atmospheres, except for TRAPPIST-1\,f and 1\,g \citep{2018NatAs...2..214D,2018AJ....156..252M};
\citet{2019AJ....157...11W} recently suggested that a clear hydrogen-dominated atmosphere may be ruled out for TRAPPIST-1\,g.
Since high-altitude clouds and haze are not expected to form in hydrogen-dominated atmospheres around temperate planets \citep[e.g.][]{2015ApJ...815..110M},
transit spectroscopy of the TRAPPIST-1 planets suggests no atmosphere or a high-metallicity atmosphere referred to as a secondary atmosphere. Future observations with precision higher than 20\,ppm is, however, needed to distinguish between the effects of cloud/haze and high metallicity in the atmospheres \citep{2018AJ....156..252M}.
Secondary atmospheres of the inner TRAPPIST-1 planets (1\,b and 1\,c) may be replenished with volcanic activity and outgassing from magma ocean due to electromagnetic induction heating \citep[][]{2017NatAs...1..878K} and tidal heating \citep{2018A&A...613A..37B}.

Atmospheric spectroscopy of the TRAPPIST-1 planets suggests that at least five of them may not retain primordial atmospheres at present. In this paper, we investigate whether a proposed formation scenario is consistent with expected atmospheric compositions of the TRAPPIST-1 planets. There are two processes that we consider in this study: accretion of the hydrogen-dominated disk gas onto a planetary core during orbital migration and atmospheric loss from a planet by photoevaporation. In Section \ref{sec:method}, we present a disk model and numerical prescriptions of gas accretion onto a migrating planetary core. In Section \ref{sec:result}, we show the amount of a hydrogen-rich atmosphere that TRAPPIST-1 like planets can acquire {\it in situ} and during orbital migration, and then estimate photoevaporative loss of the accreted hydrogen-rich gases from TRAPPIST-1-like planets.
In Section \ref{sec:discuss}, we compare our results with the atmospheric properties of the TRAPPIST-1 planets predicted by transmission spectroscopy and discuss atmospheric compositions of the TRAPPIST-1 planets and their origins. We summarize our results in the last section.

\section{Methods} \label{sec:method}

\subsection{A Migrating Planet}
We consider an isolated planetary core with mass of 0.3, 0.7, and 1\,$M_\oplus$ to be embedded in a disk. A planetary core initially resides inside/outside a snow line ($\sim$0.062\,au) around TRAPPIST-1 ($L_\star/L_\odot = 5.24 \times 10^{-4}$, where $L_\star$ is the stellar luminosity) and then accretes a disk gas {\it in situ} or during orbital migration.
Since planetary accretion proceeds rapidly in an inner region, their building blocks of planets should be almost depleted there before a planetary core starts migrating inward, although neighboring embryos may collide with each other during resonance trapping and/or orbital migration. In fact, a planetary embryo can quickly grow up to the pebble isolation mass \citep{2014A&A...572A..35L,2019A&A...627A.149S} in a high pebble-mass flux \citep{2019A&A...627A..83L}.
We assume that planetesimal/pebble accretion onto a planetary core ceases during orbital excursion,
which allows us to estimate the maximum amount of an accreting hydrogen-rich gas.

We adopt torque formulae of Type I migration, including thermal torques, on a low-mass planet in a three-dimensional and non-isothermal disk \citep{2015Natur.520...63B,2017MNRAS.471.4917J,2017MNRAS.472.4204M,2019MNRAS.486.5690G}.
The thermal torque is the sum of the cold and heating torque.
Since we assume no accretion of pebbles/planetesimals onto a planet, the heating torque comes from gas accretion processes. As shown in Section 3, the total amount of an accreted disk gas onto TRAPPIST-1-like Earth-sized planets is less than a few wt\% of their total masses. The heating torque newly included in this study does not make a significant contribution to planetary migration.
A timescale of orbital migration is almost comparable to that of radiative diffusion, i.e., gravitational contraction of a planet, in our simulations. As seen in Figure 1, the envelope evolution of a migrating planet mostly occurs after it gets stalled at a resonant location. Thus, we assume that a migrating planet can be thermodynamically equilibrated with the ambient disk gas.
In addition, the TRAPPIST-1 planets are unlikely to undergo a giant impact phase after disk dispersal as seen in the inner solar system because of resonant capture \citep{2017A&A...604A...1O}. Thus, atmospheric mass loss via giant impacts in the late stage of planet assembly can be neglected.

\subsection{Disk Model}
An initial disk model adopts the surface density profile given in \citet{2010ApJ...723.1241A}:
\begin{equation}
    \Sigma_{\rm gas} = \Sigma_0 \left( \frac{r}{R_{\rm c}} \right)^{-1} \exp{\left(-\frac{r}{R_{\rm c}} \right)},
\end{equation}
where $\Sigma_{\rm gas}$ is the surface density of a disk gas, $r$ is the distance from a central star, $R_{\rm c}$ is the characteristic disk radius, and $\Sigma_0$ is the initial surface density of a disk at $r = R_{\rm c}$, which is determined by the initial mass and size of a disk.
We adopt a radial profile of a midplane temperature in a disk with a constant $h$ used in \citet{2017A&A...604A...1O}:
\begin{equation}
    T(r) = 180 \left(\frac{M_\star}{0.08\,M_\odot} \right)
            \left(\frac{h}{0.03}\right)^2 \left( \frac{r}{0.1\,{\rm au}} \right)^{-1},
\end{equation}
where $M_\star$ is the mass of a central star and $h$ is the inner disk aspect ratio.
Disk accretion rates for M-type stars were estimated to be 10$^{-9}$--10$^{-10} M_\odot$\,{\rm yr}$^{-1}$ \citep[e.g.][]{2015A&A...579A..66M}. A midplane temperature in a disk with such a low accretion rate is little changed with time \citep[see, e.g.][]{2015A&A...575A..28B}.
Thus, we consider a fixed disk temperature profile in this study.

We consider that a disk is initially truncated at 100\,au and the initial disk mass ($M_{\rm disk}$) is proportional to the stellar mass \citep{2013ApJ...771..129A}; $M_{\rm disk}= 0.1\,M_\star$, which is gravitationally stable \citep[see Eq.(3) in][]{2016ARA&A..54..271K}. Given that the dust-to-gas ratio is 0.011 according to [Fe/H] = +0.04 of TRAPPIST-1 \citep{2016Natur.533..221G}, an initial amount of solid material of $\sim 32.6\,M_\oplus$ is available for planet formation around TRAPPIST-1 star\footnote{Since N-body simulations of TRAPPIST-1 planets predict a high efficiency of planet formation \citep{2019A&A...627A.149S}, our disk model can reconcile the total mass of the TRAPPIST-1 planets.}.
Using an empirical $M_{\rm disk}$-$R_{\rm c}$ relation\footnote{
$ \frac{M_{\rm disk}}{M_\odot} 
    \sim 2 \times 10^{-3} \left( \frac{R_{\rm c}}{10\,{\rm au}} \right)^{1.6} $}
\citep{2010ApJ...723.1241A}, we find that $R_{\rm c} \sim 25.4$\,au.

We assume that the inner edge of a disk, $r_{\rm in}$, is given by the magnetospheric cavity radius \citep{1992apa..book.....F}; \citet{2017A&A...604A...1O} estimated $r_{\rm in}\sim 0.01$\,au which is close to the current location of the innermost planet. The surface density of a disk gas declines exponetially with time, i.e., $\Sigma_{\rm gas}(t) = \Sigma_{\rm gas}(t = 0) \exp{(-t/\tau_{\rm disk})}$, where $t$ is the time and $\tau_{\rm disk}$ corresponds to the timescale of disk dispersal. We adopt $\tau_{\rm disk}=2.5$\,Myr based on disk lifetimes estimated from age--disk fraction relations in young star clusters \citep{2001ApJ...553L.153H,2008ApJ...686.1195H,2009AIPC.1158....3M,2010A&A...510A..72F,2014A&A...561A..54R}.
We consider two disk evolution models in this study: i) a steady-state, viscous accretion disk and ii) a disk wind-driven accretion disk \citep{2009ApJ...691L..49S}. 
In the former case, 
we do not simulate time evolution of the surface density of a disk gas. Instead, we simply use an exponential decay of $\Sigma_{\rm gas}(t)$.
Regarding the rapidly dissipating disk, it is pointed out that effects
of disk winds \citep{2009ApJ...691L..49S,2016A&A...596A..74S,2018A&A...615A..63O}
or photoevaporation \citep{2014prpl.conf..475A} or both mechanisms trigger a rapid
disk dispersal. For the latter case, we introduce a two-stage disk dispersal, namely, $\tau (t > \tau_{\rm disk}$)=10\,kyr.

\subsection{Gas Accretion onto a Planet}

A disk gas mainly consists of hydrogen and helium. We calculate time-dependent accretion rates of the disk gas onto a planetary core using one-dimensional hydrodynamic simulations and then estimate the mass of the hydrogen-rich atmosphere that it acquires before the disk gas disappears.
We briefly present numerical prescriptions of our hydrodynamic simulations \citep[see also][]{2012ApJ...753...66I}. A quasi-hydrostatic evolution of a planet is described by
\begin{equation}
    \frac{\partial P}{\partial M} = -\frac{GM}{4\pi r^4},
    \label{eq:P}
\end{equation}
\begin{equation}
    \frac{\partial r}{\partial M} = \frac{1}{4\pi r^2 \rho},
\end{equation}
\begin{equation}
    \frac{\partial T}{\partial M} = \frac{T}{P} \nabla \frac{\partial P}{\partial M},
\end{equation}
\begin{equation}
    \frac{\partial L}{\partial M} =  - T\frac{dS}{dt},
    \label{eq:L}
\end{equation}
where $M$, $P$, $T$, $\rho$, $L$, and $S$ are the mass enclosed within a radius, $r$, pressure, temperature, density, luminosity, and specific entropy, respectively. $\nabla$ is the temperature gradient, which is determined by heat transfer. 
The luminosity at the core surface, $L_{\rm radio}$, is the luminosity due to the radioactive decay of chondrites as core material,
$L_{\rm radio} = 2 \times 10^{20} (M_{\rm core}/M_\oplus)$\,erg\,s$^{-1}$, where $M_{\rm core}$ is the rocky core mass of a planet
\citep[see][]{1995ApJ...450..463G}.
Core cooling delays the gravitational contraction of a planet, leading to less H$_2/$He envelope mass.
We integrate Eqs.(\ref{eq:P})--(\ref{eq:L}) to determine the interior structure of a planet at a given time.
The outer boundary conditions for an accreting planet are defined by temperature and density
at the Hill radius, which are thermodynamically equilibrated with those of a disk gas.
Gravitational potential energy released by the contraction/expansion of a planetary atmosphere should compensate for heat energy which is carried away into space by radiation. Thus, we calculate the mass change of a planet that undergoes the thermal evolution.

In this study, planets are not massive enough to open a gap in the ambient disk, namely, $R_{\rm H} < H_{\rm p}$ and $M_{\rm p} < 40 \nu M_\star/(r^2\Omega)$ \citep{1993prpl.conf..749L}, where we assume an $\alpha$-disk model with a disk viscosity parameter $\alpha = 10^{-3}$ \citep{1973A&A....24..337S}, $R_{\rm H}$ is the Hill radius of a planet, $H_{\rm p}$ is the pressure scale height of a disk gas, $M_{\rm p}$ is the planetary mass, $M_\star$ is the stellar mass, $\Omega$ is the Keplerian angular velocity, $\nu$ is the kinematic viscosity given by $\nu = \alpha h^2 r^2 \Omega$, and $h$ is the disk aspect ratio. Gas accretion rates onto Earth-sized planets never exceed disk accretion $\dot{M}_{\rm disk} = 3\pi \nu \Sigma_{\rm gas}$\footnote{The supply of a disk gas to an accreting planet may be further limited in a disk wind-driven disk \citep{2018ApJ...867..127O}}. Thus, disk dispersal terminates the gas inflow toward a planet.

In this study, we assume that a disk gas freely flows into a planet until atmospheric contraction due to radiative cooling occurs.
Recently, \citet{2015MNRAS.447.3512O} found that a rapid recycle of the atmospheric gas suppresses gas accretion onto a planet embedded in an isothermal disk, whereas \citet{2018MNRAS.479..635K} showed that the atmospheric recycling of a planet should be less efficient in non-isothermal cases because of the buoyancy barrier. In addition, small grains may be suspended in an accreting H$_2$-rich gas flow, leading to the delay of atmospheric cooling \citep[][]{2017A&A...606A.146L}. 
A dusty H$_2$-rich gas, which is enriched with small grains, causes less efficient gas capture by a planet.
As a result, the final amount of a H$_2$-rich gas accreted on a planet should be reduced in a dusty (or high-metallicity) disk \citep[see also][]{2012ApJ...753...66I,2014ApJ...797...95L}. If the collisional growth of small grains in the accreted H$_2$-rich gas, however, proceeds quickly, larger grains (or aggregates) settle down and sublimate near the bottom of the planetary atmosphere \citep{2008Icar..194..368M}. Eventually, the upper atmosphere of a planet becomes grain-depleted. We consider this case in which the accreting disk gas has grain opacities
reduced to 1\% of the interstellar medium (ISM) values \citep{2003A&A...410..611S}.
Although actual grain opacities in an accreting disk gas and the upper atmosphere of a planet are still uncertain, the choice of lower opacities leads to more massive atmospheres, allowing us to estimate upper limits on the mass of an accreted hydrogen-rich gas.

\subsection{Atmospheric Escape from a Planet}

Close-in planets undergo atmospheric loss by stellar X-ray and UV (XUV) radiation and injection of high-energy particles via a stellar wind and coronal mass ejection. Atmospheric loss from the TRAPPIST-1 planets was recently studied: 
water loss from TRAPPIST-1 planets by XUV irradiation \citep[][]{2017MNRAS.464.3728B,2017AJ....154..121B} and 
atmospheric ion escape from the TRAPPIST-1 planets via a stellar wind, assuming Venus-like atmospheres \citep[][]{2018PNAS..115..260D}.
In this paper, we further estimate mass loss of a hydrogen-rich atmosphere from TRAPPIST-1-like planets (like those in the TRAPPIST-1 system) via the energy-limited hydrodynamic escape \citep[e.g.][]{1981Icar...48..150W}.

The hydrodynamic mass loss rate from a planet is given by 
\begin{equation}
    \dot{M}_{\rm esc} = \frac{\eta F_{\rm XUV} \pi R^3_{\rm XUV}}{G M_{\rm p} K_{\rm tide}},
    \label{eq:esc}
\end{equation}
where $\dot{M}_{\rm esc}$ is the mass loss rate, $F_{\rm XUV}$ is the incident XUV flux, $G$ is the gravitational constant, $R_{\rm XUV}$ is the planetary radius at which the H/He atmosphere becomes optically thick to XUV photons, $K_{\rm tide}$ is the correction factor due to the effects of stellar tidal forces \citep{2007A&A...472..329E}, and $\eta$ is the heating efficiency by stellar XUV radiation, which is defined as the ratio of kinetic energy of photoelectrons to the absorbed XUV energy. We assume that $R_{\rm XUV} \sim R_{\rm p}(t)$ for TRAPPIST-1-like planets \citep{2014ApJ...792....1L},
where $R_{\rm p} (t)$ is the planetary radius as a function of time $t$. 
We define $R_{\rm p}$ as the photosphere, i.e., $R_{\rm p} = R_{\rm bc} + R_{\rm atm}$, where $R_{\rm bc}$ is the radiative--convective boundary and $R_{\rm atm}$ is the photospheric correction given in \citet[][]{2014ApJ...792....1L}.
As outer boundary conditions for an evaporating planet after disk dissipation, it has an equilibrium temperature at the radiative--convective boundary above which is assumed to be isothermal. 
We use the equilibrium temperatures of the TRAPPIST-1 planets given in \citet{2017Natur.542..456G}.
We calculate $R_{\rm bc} (t)$ by integrating the interior structure of a planet undergoing an atmospheric mass loss at a given $t$,
using Eqs.(\ref{eq:P}--(\ref{eq:esc}).
We model the interior structure of a planet with a hydrogen-rich atmosphere using equation of state (EoS), the SCvH EoS \citep{1995ApJS...99..713S}. The core is assumed to be rocky material with a constant density in this study. We compute the thermal evolution of a planet for the age of TRAPPIST-1.
The heating efficiency for hydrogen-dominated upper atmospheres never exceeds 20\% \citep[][]{2014A&A...571A..94S,2015SoSyR..49..339I,2018MNRAS.476.5639I}. 
\citet{2012MNRAS.425.2931O} demonstrated that heating efficiencies for Earth-sized planets were low ($\eta \sim 0.1-0.15$).
Thus, we adopt a constant $\eta = 0.1$, although $\eta$ certainly varies with time.

The time evolution of the XUV flux from TRAPPIST-1 remains poorly understood. We adopt a scaling law of X-ray luminosities for M-dwarfs with ages of 5--740\,Myr found in \citet{2012MNRAS.422.2024J};
\begin{equation}
    L_{\rm XUV}(t) =
        \begin{cases}
             L_0,\,\,\,t \leq 100\,{\rm Myr} \\
             L_0 \left(\frac{t}{100\,{\rm Myr}}\right)^{-1.2},\,\,\,t > 100\,{\rm Myr}, 
        \end{cases}
\end{equation}
where $L_{\rm XUV}$ is the XUV luminosity of the star, $t$ is the age, and $L_0$ is the saturated XUV luminosity. The current XUV flux from TRAPPIST-1 was derived from {\it XMM-Newton} \citep[][]{2017MNRAS.465L..74W}, $L_{\rm XUV}/L_{\rm bol} \sim 6-9 \times 10^{-4}$, and HST/STIS observations \citep[][]{2017A&A...599L...3B,2017AJ....154..121B}, $L_{\rm XUV}/L_{\rm bol} \sim 2-4 \times 10^{-4}$, where $L_{\rm bol}$ is the bolometric luminosity of TRAPPIST-1. 

The age of TRAPPIST-1 was estimated to be $\sim 3-8$\,Gyr from the rotational period of $\sim 3.3$ days \citep{2017NatAs...1E.129L}. \citet{2017ApJ...845..110B} concluded that TRAPPIST-1 is a thin/thick disk star with an age of 7.6$\pm$2.2\,Gyr based on the analyses of Li abundance, metallicity, rotation, and the $UVW$ velocities \citep{2017ApJ...845..110B}.
Despite the old age of TRAPPIST-1, frequent strong flare events \citep[][]{2017ApJ...841..124V} and strong X-ray and EUV (XUV) emissions \citep[][]{2017MNRAS.465L..74W} from TRAPPIST-1 have been observed. The ratio of Lyman-$\alpha$ to X-ray emission from TRAPPIST-1 suggests that its chromosphere is moderately active compared to its corona and transition region \citep[][]{2017A&A...599L...3B}.
Considering that the age of TRAPPIST-1 is 7.6$\pm$2.2\,Gyr \citep{2017ApJ...845..110B},
we determine the saturated XUV luminosity of TRAPPIST-1; $L_0 \sim 4.7 \times 10^{-5} L_\odot$ for $L_{\rm XUV}/L_{\rm bol} = 5 \times 10^{-4}$.

\section{Results} \label{sec:result}

\subsection{Disk Gas Accretion {\it in situ} and During Migration}\label{sec:gas_accretion}
\begin{figure*}
\centering
\includegraphics[width=180mm, clip]{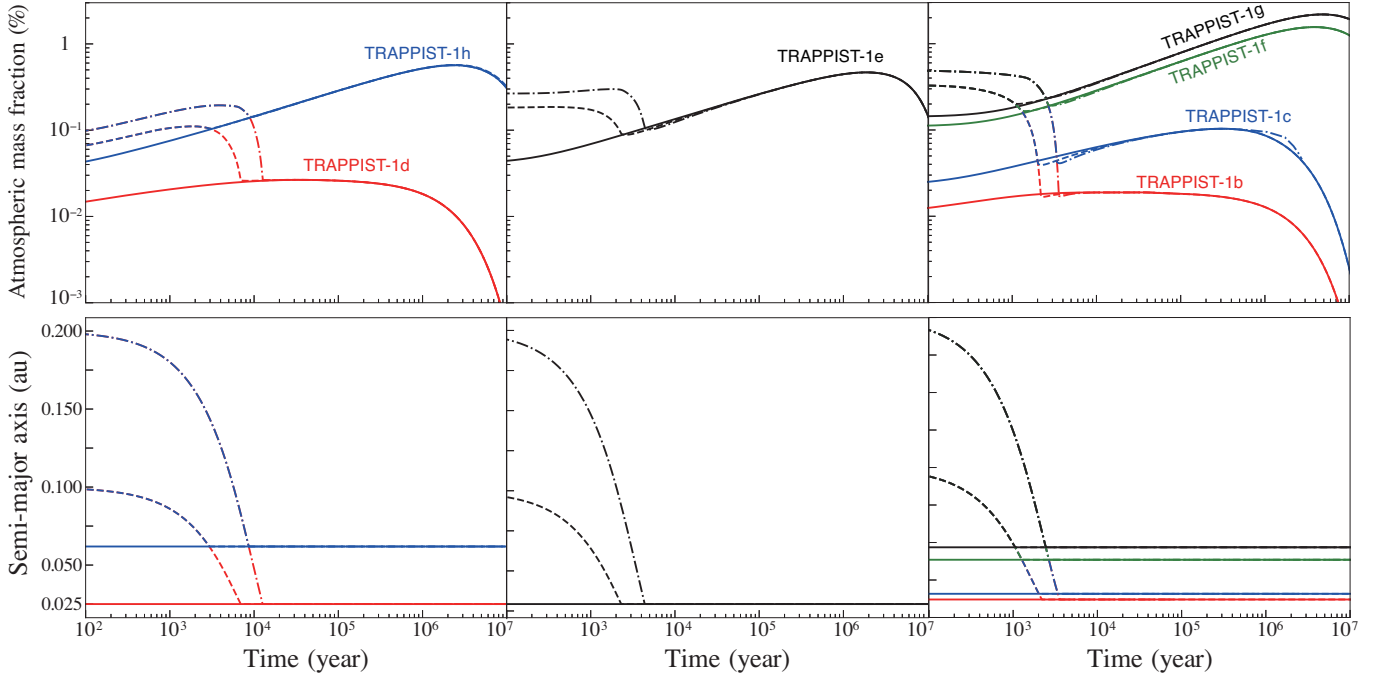}
\caption{Time evolution of the atmospheric mass fraction (top) and semimajor axis (bottom) of a non-migrating (solid curves) and migrating planetary core with mass of 0.3\,$M_\oplus$ (left: TRAPPIST-1\,d (red), 1\,h (blue)), 0.7\,$M_\oplus$ (middle: 1\,e), and 1\,$M_\oplus$ (right: 1\,b (red), 1\,c (blue), 1\,f (green), and 1\,g (black)). A planetary core starts to migrate from 0.1\,au (dashed ones) or 0.2\,au (dash-dotted ones). The locations of a non-migrating planetary core adopt semimajor axes similar to those of the TRAPPIST-1 planets. Mass loss from an accreting planet driven by a stellar XUV irradiation is not included.
}
\label{fig:ap-t}
\end{figure*}

Figure \ref{fig:ap-t} shows the atmospheric growth of a migrating and a non-migrating planetary core 
with mass of 0.3\,$M_\oplus$ (for TRAPPIST-1\,d, 1\,h), 0.7\,$M_\oplus$ (for 1\,e), and 1\,$M_\oplus$ (for 1\,b, 1\,c, 1\,f, and 1\,g)
in a steady-state accretion disk. 
The initial locations of non-migrating planetary cores adopt the orbital configuration of the TRAPPIST-1 planets,
whereas a migrating one starts to move inward from 0.1\,au or 0.2\,au.
Atmospheric loss from a planet driven by a stellar XUV irradiation is not included while gas accretion onto it proceeds.
We consider that small grains in an accreted hydrogen-rich envelope are highly depleted.

We see a stepwise atmospheric growth of a planetary core. As a migrating core approaches a central star, its atmospheric mass begins to decrease because the Hill radius shrinks and the temperature of the ambient disk gas increases. As a result, an inner planet accretes a disk gas less efficiently, as seen in Figure \ref{fig:ap-t}.
After a migrating core gets stalled at a given location, it gradually accretes the ambient disk gas {\it in situ}, leading to an upturn in the atmospheric mass. As far as the disk lifetime is longer than the Type I migration timescale, which depends on planetary mass and disk properties, the atmospheric growth of a planetary core should proceed in two phases. 

We find that disk evolution in the late stage of disk dispersal (even after $t > \tau_{\rm disk}$) controls the final atmospheric mass of a planetary core because its atmosphere continues to drain away.
A decreasing gas density at the disk midplane slows down the envelope growth and finally causes atmospheric loss during disk dissipation.
While the atmospheric mass loss occurs, the outer envelope of the planet is almost isothermal.
Since the internal energy of the planet is still slowly carried away into space, the surface temperature gradually decreases.

Atmospheric erosion of a planet in the late stage of disk dissipation is sensitive to core mass {bf and disk dissipation timescale} \citep{2012ApJ...753...66I}, core cooling \citep[][]{2016ApJ...825...29G,2018MNRAS.476..759G}, and radiative transfer in the atmosphere of a planet \citep[][]{2018MNRAS.476.2199L}. A close-in massive core, which is typically comparable to a critical core mass or larger, can avoid atmospheric loss even in a dissipating disk (see also Figure 2 of \citet{2012ApJ...753...66I}) and retain a massive atmosphere. 
Shorter disk lifetimes prevent a planet from accreting a significant amount of the disk gas. A rapid decrease in the disk gas density also drives more efficient mass loss of the envelope of a planet.
The heat released by a core drives the blow-off of thin atmospheres of TRAPPIST-1-like planets in a disk-depleted environment \citep[][]{2016ApJ...825...29G}.
In addition, since radiative cooling in the atmosphere of a planet controls the efficiency of gravitational contraction, atmospheric growth/loss should be dependent on both grain opacities and a detailed thermal profile in the atmosphere of a planet above the radiative--convective boundary \citep[][]{2018MNRAS.476.2199L}.
A massive core without a core luminosity never undergoes a significant atmospheric loss (see Figure 6 of \citet[][]{2018MNRAS.476.2199L}).

When a planet is detached from the ambient disk or becomes isolated, the remaining hydrogen-rich gas should be the gravitationally bound atmosphere. In this study, we do not simulate the detailed structure of gas flow around a planet in the last stage of disk dispersal. Nevertheless, the maximum mass fractions of an accreted hydrogen-rich atmosphere are estimated to be as small as 10$^{-2}$\,wt\% and 0.1\,wt\% for the two inner planets (TRAPPIST-1\,b and 1\,c), 10$^{-2}$\,wt\% for 1\,d,  1\,wt\%, a few\,wt\%, and a few\,wt\% for three planets (1\,e, 1\,f, and 1\,g) in a conventional habitable zone and 1\,wt\% for the outermost planet (1\,h), respectively.

 Even a $1\,M_\oplus$ core fails to acquire a massive hydrogen-rich envelope. A massive core accretes more disk gas, whereby it moves toward a central star faster than a smaller one and the Hill radius decreases. It is hard for a hot disk gas in the vicinity of a star to be gravitationally bound by a planetary core. In other words, the amount of a hydrogen-rich gas acquired by a planetary core increases with semimajor axis because of the lower thermal energy of a H$_2$/He gas and the expansion of the Hill sphere.

After orbital migration stops, the atmospheric growth of a planetary core follows in-situ accumulation of a disk gas by itself. Since a disk gas accreted onto a planetary core continues to leak out of the Hill radius with decreasing $\Sigma_{\rm gas}$, the amount of a hydrogen-rich atmosphere is controlled by atmospheric loss, accompanied by disk dissipation. The final atmospheric mass of a migrating planet is comparable to that of a non-migrating planetary core with the same mass that formed {\it in situ} after Type I migration stops. As seen in the TRAPPIST-1 system, if a massive core is captured into resonance with an inner planet during planetary migration, the amount of a hydrogen-rich atmosphere should be limited by the resonant location.

\subsection{A rapidly dissipating disk}

\begin{figure}
\centering
\includegraphics[width=0.45\textwidth]{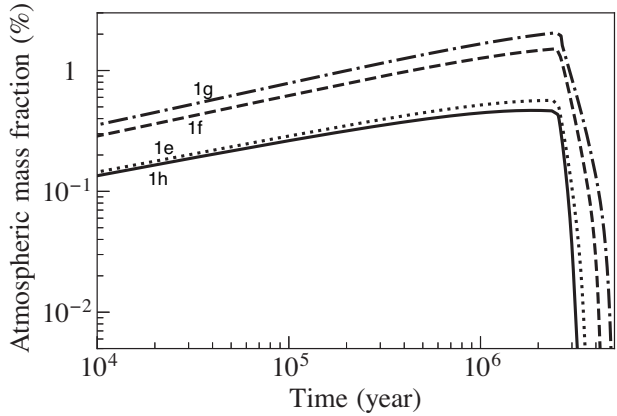}
\caption{Time evolution of the atmospheric mass fraction of a migrating planetary core with mass of 0.3\,$M_\oplus$ (solid: 1\,h), 0.7\,$M_\oplus$ (dotted: 1\,e), 1\,$M_\oplus$ (dashed: 1\,f and dashed-dotted: 1\,g) in a disk wind-driven disk.
A stellar XUV-driven atmospheric loss of an accreting planet is not included.}
\label{fig:ap-t_disk}
\end{figure}

Disk wind may accelerate disk dispersal in the late stage of planet formation \citep{2009ApJ...691L..49S}. 
We consider gas accretion onto a migrating planet in a disk wind-driven accretion disk.
Figure \ref{fig:ap-t_disk} shows atmospheric growth of a migrating planet with 0.3\,$M_\oplus$ (TRAPPIST-1\,h), 0.7\,$M_\oplus$ (1\,e), and 1\,$M_\oplus$ (1\,f and 1\,g) in a disk wind-driven accretion disk.
We see a rapid decline in the atmospheric mass of a planet after $t = \tau_{\rm disk}$ in a disk wind-driven disk. The leakage of a H$_2$-rich gas out of the Hill radius is susceptible to the decrease in the surface density of the ambient disk gas. Since rapid disk dispersal via disk wind after $t = \tau_{\rm disk}$ accelerates the decrease in density of the ambient disk gas, the amount of a H$_2$-rich gas that is gravitationally bound by a planet is suppressed by this dissipation process of a disk gas.

\subsection{Hydrodynamic escape of an accreted hydrogen-rich gas}

\begin{figure}
\centering
\includegraphics[width=0.45\textwidth]{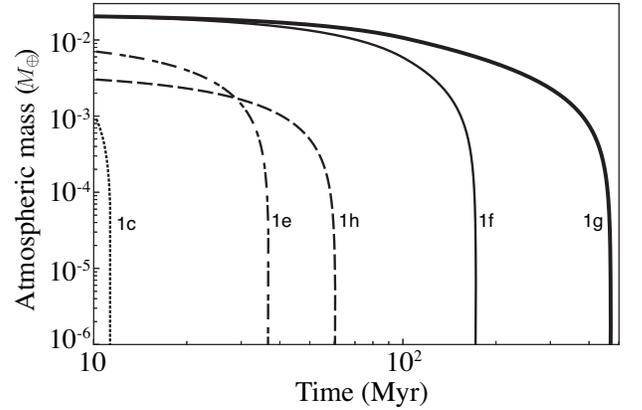}
\caption{Time evolution of the atmospheric mass of TRAPPIST-1-like planets under stellar XUV radiation. Each planet has initially the maximum amount of a hydrogen-rich atmosphere, as shown in Fig. \ref{fig:ap-t} : TRAPPIST-1\,c (0.1\,wt\% at 0.158\,au: dotted), 1\,e (1\,wt\% at 0.0293\,au: dashed-dotted), 1\,f (2\,wt\% at 0.0385\,au: thin solid), 1\,g (2\,wt\% at 0.0469\,au: thick solid), and 1\,h (1\,wt\% at 0.0619\,au: dashed). }
\label{fig:mass-loss}
\end{figure}

The TRAPPIST-1 planets can have a small amount of H$_2$-rich atmosphere ($10^{-2}$--a few wt\%) as a consequence of accumulation of a disk gas, as shown in  Figure \ref{fig:ap-t}.
Here we consider a TRAPPIST-1-like planet that retains an accreted hydrogen-rich gas after planet formation, 
and compute the atmospheric mass loss driven by a stellar XUV irradiation.
Figure \ref{fig:mass-loss} shows the atmospheric mass of only five TRAPPIST-1-like planets (1\,c: 0.1\,wt\% at 0.158\,au, 1\,e: 1\,wt\% at 0.0293\,au,
1\,f: 2\,wt\% at 0.0385\,au, 1\,g: 2\,wt\% at 0.0469\,au, and 1\,h: 1\,wt\% at 0.0619\,au) as a function of time. Since the two inner planets (1\,b and 1\,d) have a small fraction of H$_2$-rich atmosphere ($\sim 10^{-2}$\,wt\%), such atmospheres are gone in a few Myr. A smaller core closer to the central star completely loses the accreted H$_2$-rich atmosphere more rapidly. Even TRAPPIST-1-like planets in a potentially habitable zone and beyond cannot retain their primordial atmospheres for 1\,Gyr. Although time evolution of TRAPPIST-1's XUV flux remains poorly understood, 
all the primordial atmospheres of he TRAPPIST-1 planets are lost to space via hydrodynamic escape driven by a stellar XUV irradiation unless TRAPPIST-1 is a young ultracool dwarf aged of $\lesssim 1$\,Gyr.

\section{Discussions} \label{sec:discuss}

No prominent absorption features at near-infrared wavelengths in the transmission spectra of the atmospheres of the TRAPPIST-1 planets rule out cloud-free, hydrogen-rich atmospheres \citep{2016Natur.537...69D,2018NatAs...2..214D,2018AJ....156..178Z,2019MNRAS.487.1634B}, whereas a clear hydrogen-rich atmosphere for TRAPPIST-1\,f and 1\,g is still in dispute \citep{2018NatAs...2..214D,2018AJ....156..252M,2019AJ....157...11W}. Our results show that all the TRAPPIST-1 planets used to have a hydrogen-rich atmosphere of $\lesssim 10^{-2}-1$\,wt\% just after disk dispersal. 
All the accreted hydrogen-rich atmospheres of the TRAPPIST-1 planets, however, can hydrodynamically escape by a stellar X-ray and UV (XUV) irradiation from their central star in several 100\,Myr, which corresponds to the lower limit of the age of TRAPPIST-1 based on Li I absorption and the rotation period \citep[see][]{2017ApJ...845..110B}. These imply that the TRAPPIST-1 planets, including TRAPPIST-1\,g, have neither cloud-free nor cloudy/hazy, hydrogen-rich atmospheres. In other words, if the TRAPPIST-1 planets currently more or less retain atmospheres, these atmospheres likely originated from secondary processes such as volcanic activities and outgassing.

Recently, interior modeling of the TRAPPIST-1 planets predicts that they may contain water of $\lesssim$25\,wt\%
\citep{2017ApJ...842L...5Q,2018ApJ...865...20D,2018A&A...613A..68G,2018RNAAS...2b..31S,2018NatAs...2..297U,2018RNAAS...2..116U}. Since the inner planets (1\,b and 1\,c) undergo a runaway greenhouse phase, water can be easily to be lost to space.
If the initial water content of a TRAPPIST-1 planet is as high as 0.1--1\,wt\%, they can retain a significant amount of water under a strong XUV radiation field of TRAPPIST-1 \citep{2017MNRAS.464.3728B,2017AJ....154..121B}. In addition, the TRAPPIST-1 planets can survive atmospheric ion escape (O$^{+}$, O$^{+}_2$, and CO$^{+}_2$ ) driven by a stellar wind over a few 100\,Myr to $\sim$ Gyr \citep{2018PNAS..115..260D}. Thus, a Venus-like atmosphere as well water vapor might be a plausible solution to the atmospheric compositions of the TRAPPIST-1 planets, which is favored by their flat and featureless transmission spectra. If the TRAPPIST-1 planets possess terrestrial-like atmospheres containing CO$_2$, CO$_2$ absorption will be detectable in transmission spectra acquired in less than 10 transits with {\it James Webb Space Telescope}/NIRSpec Prism, although the detection of O$_2$ and O$_3$ will be still be elusive \citep{2019AJ....158...27L}.

\section{Summary} \label{sec:summary}

We have examined the accumulation of a hydrogen-rich disk gas onto a TRAPPIST-1-like planet that formed {\it in situ} and a migrating one with mass of 0.3\,$M_\oplus$, 0.7\,$M_\oplus$, and 1\,$M_\oplus$. With updated masses and semimajor axes of the TRAPPIST-1 planets, mass fractions of their hydrogen-rich atmospheres are estimated to be as small as 10$^{-2}$\,wt\% and 0.1\,wt\% for TRAPPIST-1\,b and 1\,c, 10$^{-2}$\,wt\% for 1\,d, 1\,wt\% for 1\,e, a few\,wt\% for 1\,f and 1\,g and 1\,wt\% for 1\,h, respectively.
All the accreted hydrogen-rich gases can, however, be lost to space by a stellar X-ray and UV irradiation in several 100\,Myr after disk dispersal. Our results suggest that the present-day TRAPPIST-1 planets have no primordial H$_2$-rich atmospheres. We confirm that a proposed formation scenario for the TRAPPIST-1 planets is compatible with their transit observations.
Thus, featureless transmission spectra in the atmosphere of the TRAPPIST-1 planets with HST/WFC3 imply that their atmospheres should be dominated by secondary processes such as volcanic activity and outgassing, namely, a high-metallicity gas accompanied by (no) cloud/haze. 

\acknowledgments

YH is supported by Grant-in-Aid for Scientific Research on Innovative Areas (JSPS/KAKENHI grant No. 18H05439). We thank Michiel Lambrechts for useful comments and improving our paper.

\bibliography{references}{}
\bibliographystyle{aasjournal}

\end{document}